\documentclass[a4paper,12pt]{article}
\usepackage{jheppub} 

\usepackage{graphicx}
\usepackage{dcolumn}
\usepackage{bm}
\usepackage{hyperref}

\usepackage{breqn}
\usepackage{graphicx} 
\usepackage{tikz}
\usetikzlibrary{matrix, arrows,shapes,positioning}

\newcommand{\be}{\begin{equation}}
\newcommand{\ee}{\end{equation}}
\newcommand{\ba}{\begin{eqnarray}}
\newcommand{\ea}{\end{eqnarray}}
\newcommand{\bbR}{\mathbb{R}}
\newcommand{\bbZ}{\mathbb{Z}}
\newcommand{\dd}{\text{d}}
\newcommand{\DD}{\text{D}}
\newcommand{\cM}{\mathcal{M}}
\newcommand{\cO}{\mathcal{O}}

\newcommand{\cD}{\mathcal{D}}
\newcommand{\SL}{\text{SL}}

\title{\boldmath A title with some math: $x=1$}

\abstract{In this work, we uncover a collection of non invertible topological operators linked to the 0-, 2-, 4- and 6-form symmetries related to the type IIB superstring effective theory. By pinpointing the $\SL(2,\bbZ)$-covariant conserved currents corresponding to these symmetries, we first derive a set of $\SL(2,\bbZ)$ invariant invertible topological operators that encapsulate the integer BPS charges inherent to the theory. Moving forward, by incorporating fractional charges while maintaining gauge invariance, we introduce the non invertible topological operators for each generalized symmetry, and in particular for the $\SL(2,\bbZ)$ 0-form symmetry. Identifying them as a novel kind of symmetries reminiscent of fractional quantum Hall effect (FQHE)-like non invertible operators, we study their action on charged objects and their associated SymTFTs obtained via half (higher) gauging.}

\title{On Non Invertible Symmetries in Type IIB Supergravity}

\author{Jos\'e J. Fern\'andez-Melgarejo $^1$}
\emailAdd{melgarejo@um.es}
\author{Giacomo Giorgi $^2$}
\emailAdd{giacomo.giorgi@um.es}
\author{Diego Marqu\'es $^{3,4}$}
\emailAdd{diegomarques@iafe.uba.ar}
\author{J. A. Rosabal $^1$}
\emailAdd{jarosabal80@gmail.com}
\affiliation{$^1$ Departamento de Electromagnetismo y Electr\'onica, Universidad de Murcia, Campus de Espinardo, 30100 Murcia, Spain.}
\affiliation{$^2$ Departamento de F\'isica, Universidad de Murcia, Campus de Espinardo, 30100 Murcia, Spain.}
\affiliation{ $^3$  Departamento de F\'isica, Universidad de Buenos Aires, Ciudad Universitaria, Pabell\'on 1, CABA, C1428ZAA, Argentina.}
\affiliation{$^4$ Instituto de Astronom\'ia y F\'isica del Espacio, (CONICET-UBA) Ciudad Universitaria, Pabell\'on IAFE, CABA, C1428ZAA, Argentina.}

\begin{document}
\maketitle
\flushbottom

\section{\label{sec:intro}Introduction}

In recent years, the discovery of higher-form symmetries \cite{Kapustin:2014gua,Gaiotto:2014kfa}, higher-group symmetries \cite{Sharpe:2015mja,Tachikawa:2017gyf,Cordova:2018cvg,Benini:2018reh,Cordova:2020tij}  and non-invertible symmetries \cite{Bhardwaj:2017xup,Chang:2018iay,Kaidi:2021xfk,Choi:2021kmx,Bhardwaj:2022yxj,Choi:2022zal,Bhardwaj:2022lsg,Lin:2022xod,Bartsch:2022mpm,Choi:2022jqy,Cordova:2022ieu,Bhardwaj:2022lsg}  has opened new avenues in quantum field theories (QFTs) (see \cite{Cordova:2022ruw,Shao:2023gho,Brennan:2023mmt,Bhardwaj:2023kri,Schafer-Nameki:2023jdn,Brennan:2023mmt} for some reviews). These aspects unveil new layers of complexity and richness into the symmetry structures of physical theories, shifting the paradigm of how we understand these systems.

Numerous works have significantly advanced the understanding of higher-form and non-invertible symmetries in string theory and supergravity \cite{Morrison:2020ool,Heidenreich:2021xpr,Apruzzi:2022rei,GarciaEtxebarria:2022vzq, Apruzzi:2023uma,Garcia-Valdecasas:2023mis,Bergman:2024aly,GarciaEtxebarria:2022jky,Bah:2023ymy,Lawrie:2023tdz}. See \cite{Kaidi:2024wio,Heckman:2024obe} for gauge non invertible symmetries from a worldsheet perspective.

Regarding non invertible symmetries in supergravity, we aim to understand whether new phenomena can effectively emerge at the intermediate scale between the continuous IR and discrete UV symmetries. On the other hand, dealing with both Page currents \cite{Page:1983mke} (see also \cite{Marolf:2000cb} and \cite{Maldacena:2001ss}) and higher-group structures entails an obstacle for assembling topological operators associated to these symmetries. This eventually translates to a formulation of BF theories in terms of the supergravity field strengths (see \cite{Garcia-Valdecasas:2023mis} for example).

In this paper we elaborate on this to understand the landscape of global higher-form and non-invertible symmetries in a particular supergravity theory. We want to scrutinize the symmetry structure of the bosonic sector of type IIB superstring effective theory and highlight its significance and implications, where the $\SL(2,\bbR)$ 0-form symmetry serves a crucial function.  More in detail, we would like to understand the r\^ole of the non-Abelian 0-form symmetry in the construction of the (non-)invertible topological operators of the theory and their actions.

Some works involving type IIB backgrounds in the literature amount to the obtaining of SymTFTs (symmetry topological field theories) \cite{Apruzzi:2021nmk}, the analysis of topological T-duality \cite{Zhang:2024oas}, topological duality defects \cite{Heckman:2022xgu} and branes \cite{Heckman:2022muc,Cvetic:2023plv}.

The paper is organized as follows. In Sec.~\ref{sec:IIB} we introduce the manifestly invariant $\SL(2,\bbR)$ type IIB superstring effective theory, together with its global and gauge symmetries.  In Sec.~\ref{sec:invertible} we obtain the invertible topological operators associated to the unbroken global higher form symmetries of the theory and their action on charged operators. Sec.~\ref{sec:noninv} contains a set of non invertible topological operators for each global symmetry of the theory, together with their genesis from a half higher gauging and associated SymTFTs. This, in turn, allows us to calculate their action on charged operators. Finally, our conclusions are presented in Sec.~\ref{sec:conclusions}.

\section{\label{sec:IIB} Symmetries of Type IIB supergravity}

In its manifestly  $\SL(2,\bbR)$ invariant  formulation, the bosonic type IIB supergravity  pseudoaction reads
\begin{multline}\label{iibaction}
  S  =  \int_{\Sigma_{10}} \Big[-R\star 1-\frac14 \ \dd M^{ij} \wedge\star \dd M_{ij}
   \\
   +\frac12 M^{ij}\, H_i \wedge\star H_j 
  +\frac14 \, F\wedge \star F  -\frac14 \, \epsilon^{ij}\, D\wedge H_i\wedge H_j\Big] \ ,
\end{multline}
which is complemented by the self-duality relation $F=\star F$,
where 
\be
H_i\equiv \dd B_i \ , \qquad F\equiv \dd D -\frac12 \epsilon^{ij} B_i\wedge H_j \ ,
\ee
are the field strengths that are invariant under the following gauge transformations:
\be
B_i \to B_i +\dd\lambda_i^{(1)}
\ , \qquad 
D \to D +\dd\lambda^{(3)} + \frac12 \epsilon^{ij} \dd \lambda_i^{(1)}\wedge B_j \ .
\label{eq:IIB-gauge-transf}
\ee
When $\Sigma_{10}$ has a non trivial topology that permits the existence of closed 2-forms $\Lambda_i^{(2)}$, 4-forms $\Lambda^{(4)}$, or both, the equations of motion (EOMs) are invariant under the transformations:
\be \label{2-group-B}
B_i \to B_i+  \Lambda^{(2)}_i
\ , \qquad 
D \to D +  \Lambda^{(4)}+ \frac12  \ \epsilon^{ij} \Lambda^{(2)}_i\wedge B_j \ .
\ee
Thus, while the EOMs exhibit a classical $\text{U}(1)^{(2)}\times \text{U}(1)^{(2)}$ 2-form symmetry and a $\text{U}(1)^{(4)}$ 4-form symmetry, the Chern-Simons term in the action breaks both of them for generic topologies (see also \cite{Choi:2022fgx,Garcia-Valdecasas:2023mis}). Moreover, the Bianchi identities $\dd H_i=0$ yield a $\text{U}(1)^{(6)}\times \text{U}(1)^{(6)}$ 6-form symmetry.

In contrast, this pseudoaction is  manifestly invariant under the following $\SL(2,\bbR)$ 0-form symmetry transformations \footnote{While at the level of the equations of motion (EOM), the 0-form symmetry enhances to $\bbR^+\times \SL(2,\bbR)$, in this paper we will restrict ourselves to the non-Abelian sector.}:
\be
M'_{ij}=\Omega^k{}_i\Omega^l{}_j M_{kl} \ ,  \qquad B'_i= \Omega^j{}_iB_j \ ,
\ee
where $\Omega^i{}_j\in \SL(2,\mathbb{R})$.
This $\SL(2,\bbR)$ symmetry acts as an outer automorphism on the $\text{U}(1)$ factors of both 2-form and 6-form symmetries.

Finally,  the transformations \eqref{eq:IIB-gauge-transf} allow us to identify a $(\text{U}(1)^{(2)}\times \text{U}(1)^{(2)})\times_{\kappa}\text{U}(1)^{(4)} $ 4-group classical symmetry \cite{Cordova:2018cvg}, which is fully characterized by the invariant of the $\SL(2,\bbR)$ outer automorphism, $\kappa\equiv\epsilon^{ij}$.

\section{Invertible Charged Operators and Actions}
\label{sec:invertible}

The type IIB pseudoaction is invariant only under the 0-form and 6-form symmetries. In this section we are going to obtain the invertible topological operators using their respective conserved currents.

The $\SL(2,\mathbb{R})$ codimension-1 conserved Noether current associated to the 0-form symmetry satisfies $\dd \star j_{a}=0$, where the index $a$ transforms in the adjoint representation of $\SL(2,\bbR)$ and the current $\star j_a$ is given by
\begin{multline}\label{jotanueve}
 \star j_a =  4(t_{a})^j{}_i\Big[ M^{ik}\star \dd M_{kj} + B_j\wedge M^{ik}\star H_k +\frac14\epsilon^{ik}B_j\wedge \big(B_k\wedge\star F-2 D\wedge H_k\big)\Big] \ ,
\end{multline}
where $(t_{a})^j{}_i$ are the generators of the $\SL(2,\bbR)$ group. Being this symmetry anomalous, it can be cured by adding a counterterm to the action \cite{Gaberdiel:1998ui} (see also \cite{Debray:2021vob}).

Regarding the $\text{U}(1)^{(6)}\times \text{U}(1)^{(6)}$ symmetry, the conserved current $\star j_i$ arising from the Bianchi identities of the 2-forms $B_i$ is 
\footnote{Upon dualization, such current can be recognized with the electric current associated to the shift of the 6-form dual field $\tilde B^i \to \tilde B^i+\Lambda_{(6)}^i$, with $\dd \Lambda_{(6)}^i=0$}
\footnote{Different from \cite{Garcia-Valdecasas:2023mis}, the $\SL(2,\bbR)$-covariant formulation enables us to have a doublet of invertible operators.}:
\begin{align}
    \dd \star j_i = 0 \ , \qquad \star j_i = H_i
    \ .
\end{align}

Thus, we construct the invertible topological operators $U(\Sigma_{\hat{p}})$ associated to each of the two currents as follows: 
\begin{align}
U ( \Sigma_{9})
=&\ 
\text{exp} \int_{\Sigma_{9}} \star j^{(1)} 
\ ,
&
U ( \Sigma_{3})
=&\ 
\text{exp} \int_{\Sigma_{3}} \star j^{(7)} 
\ ,
\end{align}
where $\Sigma_{\hat{p}}$ represents a $\hat{p}$-dimensional closed manifold, the currents are 
\begin{align}
\star j^{(1)} \equiv&\ q^a  \star j_a
\ ,
&
\star j^{(7)} \equiv&\ 2\pi \, \text{i} \ \tilde q^i  \star j_i
\ ,
\end{align}
and $\{q^a, \,  \tilde q^i \} $ are charges transforming in the adjoint and fundamental  $\SL(2, \bbZ)$ representations respectively. At this moment, the operators are not gauge invariant, as the charges are still real numbers.

To evaluate the action of these operators, we introduce the charged objects $\cO_{(2)}$, $\cO_{(4)}$ and $\cO_{(6)}$ which, resp., carry charges $\{Q^i,\, Q,\, \tilde Q_i\}\in\mathbb{Z}$ that correspond to F1/D1, D3 and NS5/D5 charges \cite{Marolf:2000cb}. We will denote the insertion of these operators along a generic submanifold $\Sigma_p$ as $\cO_{(p)}\equiv \cO(\Sigma_p)$ \footnote{For instance, $\cO_{(2)}$ can be understood as the $\SL(2,\bbZ)$ invariant surface operator $
    \cO_{(2)}
    \equiv 
    {\cal O} (M_2) 
    =
    \exp\oint_{M_2} i \,   Q^i B_i
    $.
}.

Then, the actions of  $U(\Sigma_{\hat p})$ on these operators are
\begin{align*}
    U(\Sigma_{\hat{p}}) \colon \begin{array}[t]{ >{\displaystyle}r >{{}}c<{{}}  >{\displaystyle}l } 
       {\cal O} (M_p) &\mapsto&  \, {\cal O} (M_p) \ .
        \end{array}
\end{align*}

Let us emphasize that the trivial action of the 0-form operator $U(\Sigma_9)$ is a consequence of the invariance of every $\cO(M_p)$ under $\SL(2,\bbZ)$. Precisely, its action on  operators $\cO_{i}(M_p)$ and $\cO_{a}(M_p)$ transforming, resp. under the fundamental and the adjoint representations is
\begin{align}
    U(\Sigma_{9}) \colon \begin{array}[t]{ >{\displaystyle}r >{{}}c<{{}}  >{\displaystyle}l } 
        \cO_{i} (M_p) &\mapsto&  \Omega^j{}_i \, \cO_{j} (M_p) 
        \ ,
        \\[.5em]
        \cO_{a} (M_p) &\mapsto&  \Omega^b{}_a \, \cO_{b} (M_p) \ ,
        \end{array}
\label{eq:inv-U9}
\end{align}
where $\Omega^j{}_i \equiv \exp{C^j{}_i }$, $\Omega^b{}_a \equiv \exp{\left(q^c f^b{}_{ca}\right)} $ and $C^j{}_i\equiv q^a (t_a)^j{}_i$.

However, for $U(\Sigma_{\hat{p}})$ to be gauge invariant, we must have $\{q^a, \, \tilde q^i \} \in \mathbb{Z}$ and this enforces the breaking of $\SL(2,\bbR)$  to $\SL(2,\bbZ)$ by quantization.

Clearly, both operators trivially account for this restriction. Nonetheless, if we still require $\Omega\in\SL(2,\bbZ)$ together with gauge invariance, then the charge matrix $C$ must be constrained. The only solution is the condition $\det C=0$. With this, $\Omega$ amounts to the $\SL(2,\bbZ)$ monodromies of the D7-brane and leaves $C$ invariant \cite{Bergshoeff:2002mb}.

For realizing the most general $\SL(2,\bbZ)$ transformation, two possibilities arise: (i) we give up the gauge invariance of $U(\Sigma_9)$ and define the action of the operator in terms of $\Omega$. This matrix can be expressed in terms of the charges $q^a\in\bbZ$ following appropriate representations as for example in \cite{Meessen:1998qm}. Or (ii) we build out an $\SL(2,\bbR)$ gauge invariant operator following the prescription of \cite{GarciaEtxebarria:2022jky} and restrict to integer values a posteriori.

\section{Non Invertibility and Action of Operators}
\label{sec:noninv}

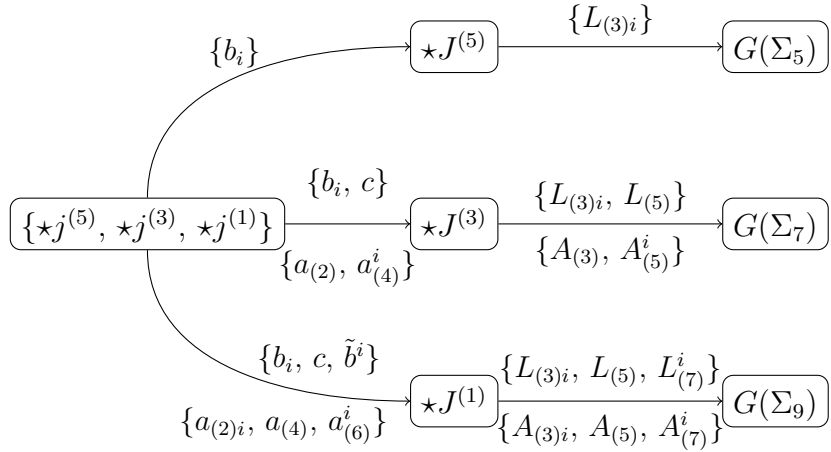
\begin{figure}
\begin{center}
\begin{tikzpicture}[scale=0.95,transform shape]
[->,>=stealth',auto,node distance=3cm,
  thick,main node/.style={circle,draw,font=\sffamily\Large\bfseries}]
  \matrix (m) [matrix of math nodes, row sep=4em, column sep=6.4em, text height=2ex, text depth=0.25ex,
  column 1/.style={nodes={draw,rounded corners}},
  column 2/.style={nodes={draw,rounded corners}},
  column 3/.style={nodes={draw,rounded corners}}
  ]
  {
   &[-10mm]   \star J^{(5)} &[3mm] G(\Sigma_5) \\
  \{\star j^{(5)}, \, \star j^{(3)}, \, \star j^{(1)}\} & \star J^{(3)} & G(\Sigma_7) \\
   & \star J^{(1)} & G(\Sigma_{9}) \\
  };

  \draw [->] (m-2-1) to [out=90,in=180]  node [above,pos=0.5] {$\{b_i\}$} (m-1-2) ;
  \draw[->] (m-1-2) -- (m-1-3) node [above,pos=0.5] {$\{L_{(3)i}\}$} ;

  \draw[->] (m-2-1) -- (m-2-2) node [above,pos=0.5,yshift=.5em] {$\{b_i,\, c\}$} node [below,pos=0.5,yshift=-.5em] {$\{a_{(2)},\, a_{(4)}^i \}$};
  \draw[->] (m-2-2) -- (m-2-3) node [above,pos=0.5] {$\{ L_{(3)i},\,L_{(5)}\}$} node [below,pos=0.5] {$\{ A_{(3)},\, A_{(5)}^i\}$};
  
  \draw [->] (m-2-1) to  [out=270,in=180] node [above,pos=0.75,yshift=.3em] {$\{b_i,\, c,\, \tilde b^i\}$} node [below,pos=0.65,yshift=-.3em] {$\{a_{(2)i},\, a_{(4)},\, a^i_{(6)}\}$} (m-3-2);
  \draw[->] (m-3-2) -- (m-3-3) node [above,pos=0.5] {$\{ L_{(3)i},\,L_{(5)},\, L_{(7)}^i\}$} node [below,pos=0.5] {$\{ A_{(3)i},\,A_{(5)},\, A_{(7)}^i\}$};
  
\end{tikzpicture}
\end{center}
    \caption{\emph{Left: Higher form currents of type IIB supergravity, where $\star j^{(7)}$ is omitted due to its trivial nature. Center: Non invertible $\star J^{(p)}$ currents and the associated auxiliary fields in the arrows. Right: Half higher gauging terms. Above and below each arrow, we show the gauge fields that induce gauge transformations on the auxiliary fields above and below the previous arrow, resp. }} 
    \label{fig:scheme}
\end{figure}

In this section we distinguish the different situations of the higher form symmetries at the level of the action.

For non trivial topologies, the $\text{U}(1)^{(2)}\times \text{U}(1)^{(2)}$ and the $\text{U}(1)^{(4)}$ symmetries are not symmetries of the pseudoaction due to the Chern-Simons terms. However, we are going to show that the theory still hosts a subgroup of them as 2-form and 4-form non invertible symmetries.

A different situation occurs with the 0- and the 6-form symmetries. Being symmetries of the action, we can promote the integers charges of their associated topological operators to rational while still preserving gauge invariance. Due to its simple structure, $\text{U}(1)^{(6)}\times \text{U}(1)^{(6)}$ does not require any further treatment for this. However, in full analogy with the FQHE, the $\SL(2,\bbZ)$ symmetry will necessarily become non invertible when considering fractionary charges.

Let us firstly introduce the classical conserved global currents associated to the 2-form and the 4-form symmetries. These, in turn,  will be improved to build gauge invariant topological operators with fractionary charges.

The conserved current of the $\text{U}(1)^{(2)}\times \text{U}(1)^{(2)}$ symmetry associated to $\Lambda_i^{(2)}$ is
\be
\dd(M^{ij}\star H_j)+\frac12 \epsilon^{ij} H_j \wedge (F+\star F) = 0 \ ,
\ee
where, upon assuming $F=\star F$, it can be rewritten as
\begin{align}
\dd\star j^i= 0,\quad \star j^i =3\Big(M^{ij} \star H_j 
+\frac{1}{3} \epsilon^{ij} B_j \wedge \big(F+2\ \dd D\big)\Big).
\label{eq:jijiji}
\end{align}

Regarding the 4-form symmetry $\text{U}(1)^{(4)}$, the conserved current  associated to $\Lambda^{(4)}$ is
\be \dd\star j= 0,\quad
\star j= 2\left(\star F+\tfrac12 \epsilon^{ij} B_i\wedge H_j\right)
\ .
\label{eq:EOM-D}
\ee

\subsection{Non invertible operators}
\label{subsec:noninv-operators}

In this section we build a set of non invertible operators associated to the generalized currents discussed above. To do so, we relax the integer condition over the charges to rational numbers, $q\, \to\,  q/N$ with $N\in\bbZ$, while preserving the gauge invariance of the operators. This is achieved by introducing a set of auxiliary fields which will be the degrees of freedom (DOFs) of the SymTFTs associated to each non invertible operator. In Fig.~\ref{fig:scheme} the arrows denote the auxiliary fields entering each non invertible operator $\cD(\Sigma_p)$.  The role of these fields is two-fold: while they allow us to restore the gauge invariance, the integration over these new DOFs makes the resulting operator non invertible. In analogy to \cite{Choi:2022jqy,Cordova:2022ieu}, our result consists of a set of FQHE operators for the symmetries of type IIB supergravity.

Let us firstly consider $U(\Sigma_3)$, the topological operator of the $\text{U}(1)^{(6)}\times \text{U}(1)^{(6)}$ symmetry. Being gauge invariant even for $q^i/N$, it does not require any treatment. Thus, there is no non invertible counterpart associated and the symmetry becomes $\bbZ^{(6)}_N\times \bbZ^{(6)}_N $.

Let us promote the charges that multiply the 0-, 2- and 4-form currents to rational values:
\be
q\, \to\, \frac{q}{N}\ , 
 \qquad  
q^i\, \to\, \frac{q^i}{N}\ , 
\qquad
 q^a\, \to\, \frac{q^a}{N}\ .
\ee
This would break the gauge invariance of some of their associated topological operators. However, upon introducing the auxiliary fields to restore the gauge invariance, they will automatically mutate to non invertible operators. 

Regarding the vector charges $q^i\in\bbZ$, and $q^a\in\bbZ$, it is sufficient considering the same integer $N$ for every component. Dividing each of its components by a different integer yields an equivalent result: $q^I/N^I= {q'}^I/N$, for $I=\{i,\, a\}$
\footnote{For example, $q^i$ can be promoted to:
$
q^i \, \to \, \Big(\frac{q^1}{N_1},\frac{q^2}{N_2}\Big) = \frac{1}{N_1N_2}(q^1N_2,\, q^2N_1)\equiv 
\frac{1}{N'}({q'}^1,\,{q'}^2)
$
}.

Let us study then the gauge invariant operator associated to the classical $\text{U}(1)^{(4)}$ that is compatible with  $q\rightarrow\frac{1}{N}q$. The Chern-Simons structure of $\star j^{(5)}$ in \eqref{eq:EOM-D} allows for the same treatment as to the original FQHE case \cite{Choi:2022jqy,Cordova:2022ieu}. In our case, the non-invertible operator results:
\begin{align}\label{O5non}
\cD(\Sigma_5)  =&\ \int [\DD b_i] \exp \Big[2\pi\, \text{i}\ \int_{\Sigma_5}  \star J^{(5)}\Big]  \ ,
\\
\star J^{(5)} \equiv&\ q \left( \frac{2}{N}\star F 
- \epsilon^{ij}\big(2H_i-N h_i\big)\wedge b_j\ \right)
\ ,
\nonumber
\end{align}
where $b_i$ is an $\SL(2,\bbZ)$ doublet of 2-forms and $h_i\equiv \dd b_i$.

We analyze now the 2-form symmetry topological operator, which is associated to $\text{U}(1)^{(2)}\times \text{U}(1)^{(2)}$, with a fractional charge $q^i\rightarrow\frac{1}{N}q^i$. In order to write a non invertible one, it is crucial to observe that the last two terms of $\star j^i$ in \eqref{eq:jijiji}, which is the non gauge invariant sector, have been expressed as the sum of two BF theories. This part can be extended  to a gauge invariant one as follows:
\begin{equation}
\label{I3}
 \epsilon^{ij} B_j \wedge \big(F+2\ \dd D\big) \quad
 \to \quad
 I^{(3)i} \equiv \epsilon^{ij}(b_j\wedge F -2 c \wedge H_j)\ ,
\end{equation}
where $b_i$ and $c$ are a doublet of 2-forms and a 4-form auxiliary fields resp., whose gauge transformations are similar to the ones in \eqref{eq:IIB-gauge-transf}:
\eqref{eq:IIB-gauge-transf}:
\begin{align*}
b_i \to&\ b_i + \dd \lambda_i^{(1)}
\ ,
&
c \to&\ c + \dd \lambda^{(3)} +\frac12 \epsilon^{ij}\dd \lambda_i^{(1)}\wedge B_j
\ .
\end{align*}
In addition, $f\equiv dc-\frac{1}{2}\epsilon^{kl}b_k\wedge H_l$ is a gauge invariant tensor mimicking the field strength $F$.

The second and final task consists of  imposing the gauge invariant constraints
\begin{align}
    H_i-Nh_i=0 \ , \qquad
    F- Nf=0 \ ,
    \label{eqbBcD}
\end{align}
whose solutions, up to gauge transformations, are
\be
\{b_i,\, c\}= \frac{1}{N}\{B_i,\, D\}
\ .
\label{solbBcD}
\ee
Thus, upon integrating out the auxiliary fields, we recover the $\bbZ_N^{(2)}\times \bbZ_N^{(2)}$ current $\star j^{(3)}/N$.
To impose the constraints \eqref{eqbBcD}, we need an extra pair of auxiliary fields,  $\{ a_{(2)},\, a^i_{(4)}\}$ which, being invariant under gauge transformations, can be understood as Lagrange multipliers. Then the operator will contain the gauge invariant pieces:
\be\label{I31}
2(F-Nf)\wedge a_{(2)}+(H_i-N h_i)\wedge a^i_{(4)} \ ,
\ee
which trivially establish the solution \eqref{solbBcD}. 
Thus, combining \eqref{I3} and \eqref{I31}, the non invertible operator is assembled as follows:
\begin{align}\label{O7non}
\cD(\Sigma_7) =& \int \DD[b_i,\, c,\, a_{(2)},\, a^i_{(4)}]\ \exp\Big[2\pi\, \text{i}\int_{\Sigma_7} \star J^{(3)} \Big]\ ,
\nonumber\\
\star J^{(3)} \equiv&\
 \frac{3}{N}q_iM^{ij} \star H_j 
 + q_i\epsilon^{ij}(b_j \wedge F -2 c \wedge H_j)\nonumber\\
& \qquad +2(F-Nf)\wedge a_{(2)}
+(H_i-N h_i)\wedge a^i_{(4)}.
\end{align}
Being gauge invariant, a non gauge invariant operator with $\bbZ_N$ charge is exactly recovered in two ways:  (i) either upon integration of the fields $\{ a_{(2)},\, a^i_{(4)}\} $, which straightforwardly implies \eqref{solbBcD}; or (ii) by integrating out the fields $\{b_i,\, c\}$. This property arises from the fact that the auxiliary fields are linearly present in each term of  \eqref{O7non}.  We emphasize that, due to the linear dependence on the auxiliary fields these integrals can be explicitly done.

Let us finally consider a non invertible operator for the $\SL(2,\bbZ)$ 0-form symmetry of the theory. 
We firstly note that, although $\star j_a$  accounts for a term proportional to $B_j\wedge M^{ik}\star H_k$, which does not admit an integration by parts, it is potentially related to the EOM of $B_i$, Eq.~\eqref{eq:jijiji}. Precisely, this EOM authorizes the introduction of a doublet of dual 7-forms, $\tilde H^i$, which is defined as 
\begin{equation}
    \tilde H^i=M^{ij} \star H_j \ ,
\label{eq:duality26}
\end{equation} and 
can be locally written as
\be\label{B6dualiza}
\tilde H^i=\dd \tilde B^i-\frac{1}{3}\epsilon^{ij} \big( B_j \wedge F- 2D\wedge H_j \big) \ ,
\ee
where $\tilde B^i$ is an $\SL(2,\bbZ)$ 6-form doublet whose gauge transformation is  
$\tilde B^i \to  \tilde B^i +\tilde \Delta^i$ with
\begin{dmath}
\tilde \Delta^i\equiv
\dd \tilde \lambda_{(5)}^i
+
\frac{2}{3}\epsilon^{ij}\dd\lambda^{(3)} \wedge B_j-\frac{1}{3}\epsilon^{ij}\left(\dd\lambda^{(1)}_j\wedge D-\frac{1}{2}\epsilon^{kl} B_j \wedge \dd\lambda^{(1)}_k \wedge B_l\right).
\end{dmath}
Based on the $\SL(2,\bbR)$-invariant democratic formulation of type IIB supergravity \cite{Fernandez-Melgarejo:2023kwk}, we introduce the auxiliary fields to construct the 0-form non invertible operator.

In particular, plugging $\tilde H^i$ into \eqref{jotanueve}, $\star j_a$ is rewritten as a composite of two BF theories:
\begin{align}\label{jotanueveB6}
 \star j_a = 4(t_{a})^j{}_i\Big[ M^{ik}\star \dd M_{kj}+\frac{1}{4}B_j\wedge\big(\tilde H^i+3\, \dd \tilde B^i\big)\Big] \ .
\end{align}
Thus, mimicking the form of $\tilde H^i$, we define the field strength $\tilde h^i$ for the 6-form auxiliary fields $\tilde b^i$ as 
\begin{equation}
    \tilde h^i 
    = 
    \dd \tilde b^i
    -\frac13 I^{(3)i}
    \ .
\end{equation}
In analogy with $\tilde B^i$, for this field strength to be gauge invariant $\tilde b^i$ transforms as $\tilde b^i \to \tilde b^i +\tilde \Delta^i$.
Therefore, the non invariant part of \eqref{jotanueveB6} is promoted to the following gauge invariant BF terms
\begin{multline}
    (t_{a})^j{}_iB_j\wedge\big(\tilde H^i+3 \dd \tilde B^i\big) \quad 
    \to \quad I_a^{(4)} \equiv (t_a)^j{}_i(b_j\wedge M^{ik}\star H_k+3B_j\wedge \dd \tilde b^i)
    \ ,
\end{multline}
where we have traded the dual 7-form $\tilde H^i$ using \eqref{eq:duality26}.  

Finally, to recover the original current we have to supplement the operator by the appropriate terms that realize the equations \eqref{eqbBcD}, together with the new gauge invariant condition
\begin{align}
M^{ij}\star H_j-N \tilde h^i=0 \ .
\label{eq:constraint-h7}
\end{align}
This is accomplished by considering the gauge invariant Lagrange-multiplier-like fields $\{a_{(2)i},\, a_{(4)},\, a^i_{(6)}\}$, which are, respectively, imposing the constraints \eqref{eqbBcD} and \eqref{eq:constraint-h7}.

Bringing together these requirements, we assemble the non invertible $\SL(2,\bbZ)$ operator
\begin{align} \label{fclO}
\cD(\Sigma_9)   =  &\ \int \DD [b_i,\, c,\, b^i,\, a_{(2)i},\, a_{(4)},\, a^i_{(6)}]\exp\Big[\int_{\Sigma_9} \star J^{(1)} \Big] 
\ ,
\\
\star J^{(1)}
 \equiv &\ \frac{4}{N}C^j{}_i M^{ik}\star \dd M_{kj}  +C^j{}_i(b_j\wedge \tilde H^i +3B_j\wedge \dd \tilde b^i)\nonumber\\
& +3(\tilde H^i-N\tilde{h}^i)\wedge a_{(2)i}  +2(F-Nf)\wedge a_{(4)}
 + (H_i-Nh_i)\wedge a^i_{(6)}\ ,
\nonumber
\end{align}
where we have used \eqref{eq:duality26}.

Similar to the previous cases, every auxiliary field appears linearly in $\star J^{(1)}$. This feature enables us to straightforwardly check that either integrating out first the fields $\{a_{(2)i},\, a_{(4)}, \, a^i_{(6)}\}$, or the set  $\{b_i,\, c, \, \tilde b^i\}$, the invertible operator $U(\Sigma_9)$ with charge $q^a/N$ is recovered.

\subsection{Half (Higher) Gaugings}
\label{subsec:higherhalf}

While half gauging a symmetry induces a codimension-1 topological interface between two different QFTs implementing a non-invertible 0-form symmetry \cite{Choi:2021kmx,Kaidi:2021xfk,Choi:2022zal}, for a higher-form symmetry the notion of higher gauging is introduced \cite{Kong:2014qka, Gaiotto:2019xmp, Roumpedakis:2022aik, Choi:2022zal, Arias-Tamargo:2022nlf,Garcia-Valdecasas:2023mis}. In this case, $p$-gauging of a $q$-form global symmetry consists of inserting a system of $q$-form symmetry defects along a codimension-$p$ manifold inside the bulk spacetime. The selfduality condition, which ensures an isomorphism between the ungauged and the gauged QFTs, is given by $q=\frac12(d+p-2)$ \cite{Choi:2022fgx}.

Additionally, $d$-dimensional theories can also be selfdual under $p$-gauging a discrete $q$-form$\times(d+p-q-2)$-form symmetry ($0\le q\le d-1$) \cite{Choi:2022fgx,Yokokura:2022alv}.

On the other hand, the obtaining of the half (higher) gauging structure provides more rigorous evidence of the topological nature of the non invertible operators \cite{Choi:2022jqy,Damia:2022bcd,Shao:2023gho}.

Apart from introducing the above auxiliary fields, this method requires some additional gauge fields to realize the gauging \cite{Choi:2022fgx,Garcia-Valdecasas:2023mis}. The arrows in Fig.~\ref{fig:scheme} show the set of necessary fields for every higher form symmetry.  Let us stress that for these gauge fields, which are just defined inside the region of the gauging, we impose Dirichlet boundary conditions $\left.A\right|_{\Sigma_p}=0$, and  $\left.L\right|_{\Sigma_p}=0$, where  $\Sigma_p=\partial \Sigma_{p+1}$  and corresponds to the closed manifold where the non invertible operator is inserted.

We also stress that the gauge fields entering the gauging procedure might be understood as the DOFs of the SymTFTs associated to each non invertible operator.

Let us firstly discuss the operator $\cD(\Sigma_5)$, which is obtained by 4-gauging the $\text{U}(1)^{(6)}\times \text{U}(1)^{(6)}$ symmetry to  $\bbZ_N^{(6)}\times \bbZ_N^{(6)}$ as follows:
\begin{align}
\label{example1}
{\cD}(\Sigma_5)
=&\
\int \DD[b_i,A_{(3)i}]\, \exp\left[\, 2\pi\, \text{i} \, G(\Sigma_5)\, \right],
\\
G(\Sigma_5)
\equiv&\
\int_{\Sigma_5}\frac{2q }{N}\star F+\int_{\Sigma_6}\frac{q}{N} \epsilon^{ij}H_i\wedge H_j -2\epsilon^{ij}A_{(3)i}\wedge(H_j-Nh_j)+N\epsilon^{ij}A_{(3)i}\wedge A_{(3)j}
\ ,
\nonumber
\end{align}
with $\Sigma_5=\partial \Sigma_6$. Here, the gauge transformation $A_{(3)i}\to A_{(3)i}+\dd\Xi_i^{(2)}$ induces the transformation $b_i\to b_i +\Xi_i^{(2)}$. In this case, $p=4$ and $q=6$, so the above self-duality condition is fulfilled. 

The operator $\cD(\Sigma_7)$ associated to $\text{U}(1)^{(2)}\times \text{U}(1)^{(2)}$ is written in terms of the half higher gauging $G(\Sigma_7)$ as:
\begin{align}\label{example2}
{\cal D}(\Sigma_7)
=&\
\int \DD[\Phi_7]\text{exp}\left[\, 2\pi\, \text{i} \, G(\Sigma_7)\, \right] \ ,
\end{align}
where $\DD[\Phi_7]\equiv \DD[b_i,\, c,\, a_{(2)},\, a^i_{(4)},L_{(3)i},L_{(5)},A_{(3)},A_{(5)}^i]$,  $\Sigma_7=\partial \Sigma_8$ is a closed manifold  and $G(\Sigma_7)$ is
\begin{align}
\label{G7}
G(\Sigma_7)=
&\ \int_{\Sigma_7} \frac{3q_i}{N}M^{ij}\star H_j\\
&\ +\int_{\Sigma_8} \frac{3q_i}{N}\epsilon^{ij}H_j\wedge F  +(H_j-Nh_j)\wedge A_{(5)}^j\nonumber\\
&\ +L_{(3)j}\wedge (-q_i \epsilon^{ij}F-Nda^j_{(4)}-N\epsilon^{ji}H_i\wedge a_{(2)})  -N L_{(3)j}\wedge A_{(5)}^j\nonumber\\
&\ +2(F-Nf)\wedge A_{(3)}  +2L_{(5)}\wedge(q_i\epsilon^{ij}H_j-Nda_{(2)})- 2 N L_{(5)}\wedge A_{(3)}\nonumber .
\end{align}
Gauge invariance under transformations of the new fields $\{L_{(3)i},L_{(5)},A_{(3)},A_{(5)}^i\}$ yields nontrivial gauge transformations of $\{b_i,\, c,\, a_{(2)},\, a^i_{(4)}\}$. As $p=2$, $q=4$, the self duality amounts to a discrete $\bbZ_N^{(4)}\times \bbZ_N^{(6)}$ symmetry \cite{Choi:2022fgx}.

Finally, we consider  the codimension-1 defect $\cD(\Sigma_9)$ associated to the $\SL(2,\bbR)$ 0-form symmetry. Here the operator is written in terms of $G(\Sigma_9)$ as:
\begin{align}\label{example3}
{\cD}(\Sigma_9)
=&\
\int \DD[\Phi_9]\text{exp}\left[ G(\Sigma_9)\right]
\ ,
\end{align}
where $\DD[\Phi_9]\equiv \DD[b_i,\, c,\, \tilde b^i,\, a^{(2)}_i,\, a_{(4)} ,\, a^i_{(6)} ,\, L^{(3)}_i ,\, L^{(5)} ,\, L^i_{(7)} ,\, \\
A^{(3)}_i ,\, A^{(5)} ,\, A^i_{(7)}]$,  $\Sigma_9=\partial \Sigma_{10}$ is closed and $G(\Sigma_9)$ is
\begin{align}
\label{G9}
G(\Sigma_9)
=&\ \int_{\Sigma_9} \frac{4}{N} C^j{}_i M^{ik} \star \dd M_{kj} \\
\hspace*{-2cm}&\ +\int_{\Sigma_{10}}
 \frac1N C^j{}_i  \left(H_i\wedge \tilde{H}^i-\epsilon^{ik}B_j\wedge H_k\wedge F\right)\nonumber \\
&\ -NL_i^{(3)}\wedge\Big(
\frac{C^i{}_j}{N} \tilde{H}^j+\dd a_{(6)}^i-\epsilon^{ij} a_j^{(2)}\wedge F 
-\epsilon^{ij} H_j\wedge a_{(4)} \Big)  + (H_i-Nh_i)\wedge A_{(7)}^i
\nonumber\\ 
&\ -NL_i^{(3)}\wedge A_{(7)}^i \nonumber \\
&\ +2NL_{(5)} \wedge\left(-N \dd a_{(4)}+\epsilon^{ij} a_i^{(2)}\wedge H_j \right)  +2(F-Nf)\wedge A_{(5)}-2NL_{(5)}\wedge A_{(5)}\nonumber\\
&\ +3L_i^{(7)}\wedge \left(-N \dd a_i^{(2)}+ C^j{}_i H_j \right)  +3(\tilde{H}^i - N\tilde h^i)\wedge  A_i^{(3)}-3NL^i_{(7)}\wedge A_i^{(3)}\ .\nonumber 
\end{align}

Interestingly ($p=0,\, q=2$), the 2-form and 6-form symmetries are democratically gauged to $(\bbZ_N^{(2)}\times \bbZ_N^{(2)})\times (\bbZ_N^{(6)}\times \bbZ_N^{(6)})$. However, as expected from its invariance under $\SL(2,\bbR)$, the 4-form symmetry is fictitiously gauged. This is also confirmed in the next section.

Thus, the discovery of these non invertible operators implies the existence of potentially new SymTFTs that are described by the actions $G(\Sigma_p)$ in \eqref{example1}, \eqref{G7} and \eqref{G9}.

\subsection{Action of Operators}
\label{subsec:noninv-action}

The half higher gauging method has been proven to be useful to calculate the action of the noninvertible operators on charged objects \cite{Chang:2018iay,Choi:2022jqy,Choi:2022fgx}.

Starting with $\cD(\Sigma_3)\equiv U(\Sigma_3)$ with $\tilde q^i\to 
\tilde q^i/N$, its nontrivial and invertible action on the aforementioned charged operators reduces to:
\begin{align}
    \cD(\Sigma_3) \colon \begin{array}[t]{ >{\displaystyle}r >{{}}c<{{}}  >{\displaystyle}l } 
        {\cal O} (M_6) &\mapsto& e^{2\pi \text{i}\, \frac{\tilde q^i \tilde  Q_i}{N}} \, {\cal O} (M_6)
        \ .
        \end{array}
\end{align}

The action of the 4-form symmetry operator $\cD(\Sigma_5)$ is
\begin{align}
    \cD(\Sigma_5) \colon \begin{array}[t]{ >{\displaystyle}r >{{}}c<{{}}  >{\displaystyle}l } 
        {\cal O} (M_2) &\mapsto&  {\cal O} (M_2) \ ,
        \\[.5em]
        {\cal O} (M_4) &\mapsto& e^{2\pi \text{i}\, \frac{q Q}{N}} \, {\cal O} (M_4) \ ,
        \\[.5em]
        {\cal O} (M_6) &\mapsto&  {\cal O} (M_6) \, e^{-2\pi\text{i}\,\tilde Q_i\epsilon^{ij}\int_{\cM_3}2A_{(3)j}} \ ,
        \end{array}
\end{align}
where $\cM_3\subset M_6\cap \Sigma_5$ is a compact manifold such that $\oint_{\partial\cM_{3}} \Xi_{(2)i}\in2\pi\bbZ_i$. Using the EOM of $A_{(3)i}$ we obtain
\begin{align}
    \cD(\Sigma_5) \,
        {\cal O} (M_6) =  {\cal O} (M_6) \,  e^{-2\pi\text{i}\, \epsilon^{ij}\frac{2\tilde Q_i}{N}\int_{\cM_3}H_{j}}\ .
\end{align}

Secondly, the action of $\cD(\Sigma_7)$ amounts to
\begin{align*}
     \cD(\Sigma_7) \colon \begin{array}[t]{ >{\displaystyle}r >{{}}c<{{}}  >{\displaystyle}l } 
        {\cal O} (M_2) &\mapsto& e^{2\pi\text{i}\, \frac{3q_i Q^i}{N}}\,  {\cal O} (M_2)
        \ ,
        \\[.5em]
        {\cal O} (M_4) &\mapsto& \,  {\cal O} (M_4) e^{2\pi\text{i}\, Q \int_{\cM_{3}} 2A_{(3)}+ q_i\epsilon^{ij} L_{(3)j}} \ ,
        \\[.5em]
        {\cal O} (M_6) &\mapsto& {\cal O} (M_6) \,  e^{2\pi\text{i}\, \tilde Q_i\int_{\cM_5}A^i_{(5)}+2\epsilon^{ij}q_j L_{(5)}}
        \ ,
        \end{array}
\end{align*}
where $\cM_3\subset M_4\cap \Sigma_7$ and $\cM_5\subset M_6\cap \Sigma_7$ are submanifolds such that the integration of the gauge parameters of $\{L_{(3)i} ,\, A_{(3)} \}$ and $\{L_{(5)},\, A_{(5)}^i\}$ are, resp., quantized over $\partial\cM_3$ and $\partial\cM_5$. Subsequently, if we evaluate the non invertible phases onshell, we obtain:
\begin{align}
     \cD(\Sigma_7) \, {\cal O} (M_4) =&\   {\cal O} (M_4) \, e^{2\pi\text{i} \, Q \frac{5q_i \epsilon^{ij}}{N}\int_{\cM_{3}} H_j} \ ,
        \\[.5em]
      \cD(\Sigma_7) \,   {\cal O} (M_6) =&\  {\cal O} (M_6) \, e^{2\pi\text{i}\,  \tilde Q_i \frac{5 \epsilon^{ij}q_j}{N}\int_{\cM_5 } F} 
        \ .
\end{align}

Finally, the action of the $\SL(2,\bbR)$ non invertible operator nontrivially acts over all the charged objects:
\begin{align*}
    \cD(\Sigma_9) \colon \begin{array}[t]{ >{\displaystyle}r >{{}}c<{{}}  >{\displaystyle}l } 
        {\cal O} (M_2) &\mapsto&   {\cal O} (M_2)\, e^{-Q^i \int_{\cM_3} L^{(3)}_i+C^j{}_i A^{(3)}_j}
        \ ,
        \\[.5em]
        {\cal O} (M_4) &\mapsto&    {\cal O} (M_4)\, e^{-Q\int_{\cM_5} L_{(5)}}
        \ ,
        \\[.5em]
        {\cal O} (M_6) &\mapsto&   {\cal O} (M_6)\, e^{ \tilde Q_i \int_{\cM_7} 3 C^i{}_j A_{(7)}^j -L_{(7)}^i }
        \ ,
        \end{array}
\end{align*}
such that $\partial\cM_3= M_2\subset \Sigma_9$, $\partial\cM_5= M_4\subset \Sigma_9$ and $\partial \cM_7= M_6\subset \Sigma_9$. Using the EOMs of the gauge fields $L$s and $A$s, these non invertible phases result:
\begin{align}
    \cD(\Sigma_9) \, {\cal O} (M_2) 
    =&\   {\cal O} (M_2) \, e^{Q^i\frac{4}{N}C^j{}_i \int_{\cM_3} H_j}
        \ ,
        \\[.5em]
    \cD(\Sigma_9) \, {\cal O} (M_4) =&\    {\cal O} (M_4)
        \ ,
        \\[.5em]
    \cD(\Sigma_9) \, {\cal O} (M_6) =&\   {\cal O} (M_6) \, e^{- \tilde Q_i \frac{4}{N}C^i{}_j\int_{\cM_7} M^{jk}\star H_k }
        \ .
\end{align}

\section{Conclusions}
\label{sec:conclusions}

In this work we have studied the higher form symmetries of type IIB supergravity and found the invertible topological operators associated to them. Interestingly, we have obtained an invertible operator that carries out $\SL(2,\bbZ)$ transformations on operators or objects charged under this symmetry. We have provided a mechanism to systematically obtain a set of noninvertible operators inspired by BF theories. Then, led by the half higher gauging construction, we have obtained their action on charged objects.  Being this a first step, a more detailed analysis of these operators \cite{Freed:2006yc} could shed some light on their uniqueness and potential applications. 

Likewise, understanding the half higher gauging as an effective description of some phenomena occurring at intermediate energy scales in supergravity  (with generalized symmetries in gravity \cite{Gomez-Fayren:2024cpl}) could be interesting.

Finally, we would like to understand whether there exists a relation between the operators with continuous symmetries and the fluxbranes discussed in \cite{Cvetic:2023plv}. It would be interesting to compare the SymTFTs and their associated DOFs.

\begin{acknowledgments}

 We thank Federico Bonetti, I\~naki Garc\'ia-Etxebarr\'ia, Miguel Montero, Tom\'as Ort\'in and Luca Romano for useful discussions and comments. D.M. thanks Universidad de Murcia for hospitality. D.M. is supported by Consejo Nacional de Investigaciones Cient\'ificas y T\'ecnicas (CONICET) and Universidad de Buenos Aires (UBA). The work of J.J.F.-M., G.G. and J.A.R. has been supported in part by the MCI, AEI, FEDER (UE) grant PID2021-125700NAC22. The work of G.G. has been supported by the predoctoral fellowship FPI-UM R-1006-2021-01. J.J.F.M. acknowledges ``El poder del arte'' for support. 

\end{acknowledgments}

\newpage

\bibliographystyle{JHEP}
\bibliography{apssamp.bib}

\end{document}